Updated guidelines, updated curriculum: The *GAISE College Report* and introductory statistics for the modern student

Beverly L. Wood, Megan Mocko, Michelle Everson, Nicholas J. Horton, and Paul Velleman

Since the 2005 American Statistical Association's (ASA) endorsement of the *Guidelines for Assessment and Instruction in Statistics Education College Report*, changes in the statistics field and statistics education have had a major impact on the teaching and learning of statistics. We now live in a world where "Statistics – the science of learning from data – is the fastest-growing science, technology, engineering, and math (STEM) undergraduate degree in the United States," [according to the ASA,](#) and where many jobs demand an understanding of how to explore and make sense of data. Some other changes include:

- More students study statistics. Often exposure to statistical thinking begins in grades 6 through 8 as numerous states adopt the [Common Core State Standards](#) or similar sets of educational standards for mathematics. Other states might expose students to statistical ideas even earlier, as part of the elementary school mathematics curriculum. Although the GAISE PreK-12 report, developed by Christine Franklin and colleagues, presents a framework for teaching statistics at the pre-college level, the GAISE College Report recommendations can still be introduced at this level and then enhanced at the post-secondary level. This might be especially useful in light of the many students who will become future elementary and secondary mathematics and statistics teachers.
- The growth in available data has made the field of statistics more salient and has provided rich opportunities to address important statistical questions even in an introductory course.
- The discipline of "Data Science" has emerged as a field that encompasses elements of statistics, computer science, and domain-specific knowledge.
- More powerful and affordable technology options have become widely available.
- Alternative learning environments have become more common.
- Innovative ways to teach the logic of statistical inference have received increasing attention.

In addition, a number of reports endorsed by the ASA – and available at [www.amstat.org/asa/education/undergraduate-educators.aspx](http://www.amstat.org/asa/education/undergraduate-educators.aspx) – have addressed aspects of statistics education:

- *Curriculum guidelines for undergraduate programs in statistical sciences*

- *Statistical education of teachers*
- *Qualifications for teaching introductory statistics*

The ASA's statement on *p*-values in *The American Statistician* also addresses the teaching of a core concept in an introductory course.

In light of these new reports and other changes and demands on the discipline, a group of volunteers led by Michelle Everson (Ohio State University) and Megan Mocko (University of Florida) have revised the 2005 GAISE College Report. The updated report was endorsed by the Board of Directors of the American Statistical Association in July 2016.

The Executive Summary of the recently endorsed *Guidelines for Assessment and Instruction in Statistics Education College Report 2016* (reprinted in full in Table 1) motivates why the original report needed revision and summarizes key changes in the new report.

Table 1. Executive Summary—Guidelines for Assessment and Instruction in Statistics Education College Report 2016 (endorsed by the ASA Board of Directors, July 2016)

> In 2005, the American Statistical Association (ASA) endorsed the *Guidelines for Assessment and Instruction in Statistics Education* (GAISE) College Report. This report has had a profound impact on the teaching of introductory statistics in two- and four-year institutions, and the six recommendations put forward in the report have stood the test of time. Much has happened within the statistics education community and beyond in the intervening 10 years, making it critical to re-evaluate and update this important report.
> 
> The revised GAISE College Report takes into account the many changes in the world of statistics education and statistical practice since 2005 and suggests a direction for the future of introductory statistics courses. Our work has been informed by outreach to the statistics education community and by reference to the statistics education literature.
> 
> We continue to endorse the six recommendations outlined in the original GAISE College

> Report.  We have simplified the language within some of these recommendations and re-ordered other recommendations so as to focus first on *what* to teach in introductory courses and then on *how* to teach those courses.  We have also added two new emphases to the first recommendation.  The revised recommendations are:
>
> 1. Teach statistical thinking.
>    - Teach statistics as an investigative process of problem-solving and decision-making.
>    - Give students experience with multivariable thinking.
> 2. Focus on conceptual understanding.
> 3. Integrate real data with a context and purpose.
> 4. Foster active learning.
> 5. Use technology to explore concepts and analyze data.
> 6. Use assessments to improve and evaluate student learning.
>
> The revised GAISE report includes an updated list of learning objectives for students in introductory courses, along with suggested topics that might be omitted from or de-emphasized in an introductory course. In response to feedback from statistics educators, we have substantially expanded and updated some appendices. New appendices provide a history of the evolution of introductory statistics courses, examples involving multivariable thinking, and suggestions for ways to implement the recommendations in a variety of different learning environments.

To help shed additional light on the revision process and subsequent changes in the report, we will share insights into the committee's thoughts and assumptions.  Next, we will briefly describe each of the key recommendations, then close with discussion of future work and next steps.

**Insight**

Throughout the revision process, the committee remained cognizant of the tension between realistic expectations and aspirational goals, and the need to be more descriptive rather than prescriptive.  There is dramatic variability in how introductory statistics is taught and the curriculum that is delivered, and that variability is increasing. We acknowledge the multiple constraints instructors face when attempting to change their courses and curricula.  While some instructors and programs may enjoy the freedom to completely restructure their courses from scratch, others may find that regular incremental review and change combined with periodic assessment is more feasible. The revised report supports either approach via a set of augmented appendices that include numerous examples of activities, assessments, and technology tools, which can be used in a variety of learning environments and settings. The GAISE guidelines can serve as an important guide to make sure that the course changes remain on track with future trends in the discipline of statistics.

The revised report acknowledges that there is not a single introductory statistics course but a variety of courses reflecting the needs of particular institutions and particular students. Some of this variability is introduced by whether a course is discipline-specific or general education, whether it is a seminar, large lecture, or a distance class, or whether algebra or calculus is a prerequisite. We believe, as did our predecessors who wrote the original GAISE College Report, that the GAISE College Report 2016 can usefully inform the pedagogy of any instructor. (In fact, the committee asserts that these fundamental ideas are also relevant for courses **beyond** the introductory course.)

Like the original report, the revision includes sections on *Goals for Students* and *Recommendations for Teaching*. The goals have been reorganized but retain non-prescriptive advice while providing structure relevant for a wide variety of courses. The committee expanded the list to include remarks about data ethics and statistical software to acknowledge inclusion of these areas in courses with a focus on contemporary statistical practice. Further, as a way of acknowledging the realities of limited time in the curriculum, there is also a section in the report for instructors addressing how they might make room by looking at a list of *Suggestions for Topics that Might be Omitted*. In an effort to help instructors meet the recommendations regardless of their learning environments, Appendix F *Learning Environments* was added with methods for meeting the recommendations in diverse settings.

## Key Recommendations

One could argue that *little* has changed in terms of the recommendations from the revised GAISE College Report compared to the 2005 Report developed by Joan Garfield and colleagues.

It would also be reasonable to posit that *everything* has changed given the dramatic shift in landscape for statistics and data science. The truth is somewhere in between.

We believe that the six recommendations of the original report have held up well over time, so these have essentially stayed the same with small changes in wording and order. These recommendations appear in the revised report with additional resources from recent statistics education literature and specific *Suggestions for Teachers* with frequent references to examples in the appendices. The primary change has been the addition of two new emphases under the recommendation to teach statistical thinking, and, therefore, the two new emphases have extra details. The first of these emphases (Teach statistics as an investigative process of problem-solving and decision-making) begins with the statement, "Students should not leave their introductory statistics course with the mistaken impression that statistics consists of an unrelated collection of formulas and methods," and the second emphasis (Give students experience with multivariable thinking) includes, "We must prepare our students to answer challenging questions that require them to investigate and explore relationships among many variables."

In terms of the first emphasis, some textbook authors have already begun to take a more holistic approach to statistics in the introductory – and potentially only – course in statistics. The emphasis on this aspect of teaching statistical thinking is meant to inspire more authors to do the same and to encourage instructors to be cognizant of the big picture perspective that may be missing from their current course. For example, rather than teach statistics as a list of topics that students might perceive as being disconnected from one another, important content can be taught from the first day of class and repeated frequently through case studies or engaging examples that highlight the investigative nature of the discipline. Students can learn the process of beginning with a research question and crafting a plan to gather and explore relevant data in order to draw appropriate conclusions about the research question. Instructors should be encouraged to think about how to incorporate this approach into their current curriculum.

For the second emphasis, a stand-alone appendix on multivariable thinking provides specific examples of extensions from more typical content in an introductory course that can help instructors add some exposure without adding too much additional content to an already crowded syllabus. Note that this does not necessarily require the use of advanced inferential techniques, but rather encourages the students to explore beyond two variables, perhaps with visualization tools, stratification, or the use of multiple regression as a descriptive approach. For example, if students are already exploring the relationship between two quantitative variables such as gas mileage and weight of a car in class, the students might have ideas about what other factors might affect these variables. They might wonder or the instructor could posit, "Is there a difference if the car is made domestically or not?" This additional information can be explored by creating scatterplots for each group or overlaying them in some fashion.

We will now give a brief summary of each of the six recommendations, along with some motivating examples and elaborations.

**Recommendation 1: Teach statistical thinking**

An introductory course is often the only statistics course that a student takes. As such, it is important that we think carefully about what our focus should be in this course: what should we teach, and what skills do we want our students to have when they leave the course? Will students use statistics in follow-up courses and careers, and will they be consumers of statistical information presented in the news and abounding in everyday life?

It is essential to work on the development of skills that will allow students to think critically about statistical issues and recognize the need for data, the importance of data production, the omnipresence of variability, and the quantification and explanation of variability. As part of the development of statistical thinking skills, it is crucial to focus on helping students become better educated consumers of statistical information by introducing them to the basic language and the fundamental ideas of statistics, and by emphasizing the use and interpretation of statistics in everyday life.

We urge instructors of statistics to emphasize practical problem-solving skills that are necessary to answer statistical questions. We should model statistical thinking and the statistical analysis process for our students throughout the course, rather than present students with a set of isolated tools, skills, and procedures. Effective statistical thinking requires seeing connections among statistical ideas and recognizing that most statistical questions can be solved with a variety of procedures and that there is often more than one acceptable solution. Instructors can illustrate statistical thinking by working through examples for students and explaining, along the way, the questions that come up and the processes involved in taking a problem from its conception to its conclusion. For example, the instructor can ask questions such as "What is the question being asked?", "Why do we need data to answer this question?", "What about variation?", "How do we make decisions based on the data?", "How do we account for variability in that decision?" and "Are there limitations on our conclusions based on how our data were measured and/or collected?" Wild and Pfannkuch give a detailed breakdown of statistical thinking in their 1999 paper.

We should teach students that the practical operation of statistics is to collect and analyze data to answer questions. As a part of the overarching recommendation to teach statistical thinking, we propose that the introductory course teach statistics as an investigative process of problem-solving and decision-making. We also propose that all students be given experience with multivariable thinking in the introductory course. Rather than limiting the focus to single variables or relationships between two variables, students should be presented with examples that illustrate how multiple variables relate to one another. This is especially important because the types of questions students are likely to encounter beyond the introductory course—perhaps within their own fields of study—could easily involve the interplay among many variables. An example that is described in more detail in the report involves average teacher salary and average SAT score across the 50 states and how the relationship between these variables—and the resulting interpretation—changes when the proportion of students who take the SAT in different states is taken into account.

**Recommendation 2: Focus on conceptual understanding.**
Students need to be able to make decisions about the most appropriate ways to visualize, explore, and, ultimately, analyze a set of data. It will not be helpful for students to know about the tools and procedures that can be used to analyze data if they don't first understand the underlying concepts. Procedural steps too often claim students' attention. Teachers should instead direct their attention toward concepts. When students have a sound understanding of concepts such as variability, bias, randomness, distribution, and inference they find it easier to use necessary tools and procedures to answer particular questions about a dataset. Students with a good conceptual foundation from an introductory course will be better prepared to study additional statistical techniques in a second course.

**Recommendation 3: Integrate real data with a context and a purpose.**

It is crucial to use real data in context for teaching and learning statistics, both to give students experience with analyzing genuine data and to illustrate the usefulness and fascination of our discipline. Statistics can be thought of as the science of learning from data, so the context of the data becomes an integral part of the problem-solving experience. The introduction of a data set should include a context that explains how and why the data were produced or collected. Students should practice formulating strong questions and answering them appropriately based on how the data were produced and analyzed. Using real data sets of interest to students is an effective way to engage students in thinking about data and relevant statistical concepts. One way this could be accomplished is to gather data from students, either as part of a class activity or through a survey administered during the start of the course. This class-generated data could then be used to investigate relationships among variables or to illustrate different ways to graph and explore particular types of variables. Other examples of real data can be found within online open data resources or perhaps through collaboration with a local researcher.

**Recommendation 4: Foster active learning.**

Students play an active role in the learning process when then they take responsibility for their learning. This might involve, for example, collaborating with peers by discussing or debating concepts and ideas, teaching their peers, or working through problem-solving exercises. Active learning methods allow students to discover, construct, and understand important statistical ideas, and engage in statistical thinking. Other benefits include the practice students get communicating in statistical language and learning to work in teams to solve problems. Activities provide teachers with a method of assessing student learning and provide feedback to the instructor on how well students are learning and shift the classroom role of the instructor from lecturing to facilitating. Instructors should not underestimate the learning gains that can be achieved with activities or overestimate the value of lectures to convey information. Embedding even brief activities within lectures can break the natural occasional dips in attention experienced by passive or minimally-engaged listeners.

Some rich activities can take an entire class session, but many valuable activities need not take much time. One active learning technique is known as *think-pair-share*. Students are presented with a question, asked to think about a response to the question and then asked to discuss that response with a neighbor. The instructor might then call upon different pairs of students to share their responses with the whole class. A *think-pair-share* discussion or prediction exercise may take only two or three minutes. Collecting on-the-spot data may take more time but reaps benefits beyond the single activity that prompted the collection. Technology can also be used through a course management system or an audience response system to decrease the time required.

**Recommendation 5: Use technology to explore concepts and analyze data.**

Technology has changed the practice of statistics and hence should change what and how we teach. By technologies, we refer to a range of hardware and software that can do far more than handle the computational burden of analysis. By adopting the best available tools (subject to institutional constraints), we allow students to *do* analysis more easily and therefore open up time to focus on interpretation of results and testing of conditions, rather than on computational mechanics. Technology should aid students in learning to think statistically and to discover concepts, sometimes through the use of simulation. It should also facilitate access to real (and possibly large) datasets, foster active learning, and embed assessment into course activities.

Statistics is practiced with computers, usually with specially designed computer software. Students should learn to use a statistical software package if possible. Calculators can provide some limited functionality for smaller datasets, but their use should be supplemented with experience reading typical output from statistical software (e.g. Minitab, SPSS, R/RStudio,…) results. Regardless of the tools used, it is important to view the use of technology not just as a way to generate statistical output but as a way to explore conceptual ideas and enhance student learning.

**Recommendation 6: Use assessments to improve and evaluate student learning.**

Students will value what you assess; therefore, assessments need to be aligned with learning goals. Assessments need to focus on understanding key ideas, and not just on skills, procedures, and computed answers. Being able to calculate a *p*-value is not enough; students need to be able to draw conclusions about the research question from a *p*-value and also explain the reasoning process that leads from the *p*-value to the conclusion.

Useful and timely feedback is essential for assessments to lead to learning. We encourage instructors to maximize opportunities to include formative assessments into their courses rather than focus exclusively on summative assessments. We also encourage instructors to review model assessments created by key Statistics Education leaders, such as the Advanced Placement Statistics Exam, the Comprehensive Assessment of Outcomes in Statistics (CAOS), and the Level of Conceptual Understanding in Statistics (LOCUS).

## Closing Thoughts

The six Recommendations for Teaching are essentially unchanged from the original GAISE report but have more robust explanations and discussion; in particular, they are supported by references to statistics education research literature, much of which did not exist in 2005. The two new emphases given to the recommendation "Teach statistical thinking" may strike some readers as being among the aspirational aspects of the new report, though they are supported by the statistics education community discourse over the previous decade. Intentional focus on the investigative nature of statistical study which leads to problem-solving and decision-making, as well as acknowledgement that many of those problems and decisions require the handling of multivariable data, bring the classroom experience closer to modern statistical practice. To find out more about what is written in the original and revised reports go to www.amstat.org/asa/education/Guidelines-for-Assessment-and-Instruction-in-Statistics-Education-Reports.aspx .

There are many challenges inherent in changing curricula. In his closing rejoinder to the discussion of his provocative paper in the *American Statistician*, George Cobb stated that our thinking about the statistics curriculum needs to start from the ground up, "starting necessarily with alternatives to the former consensus introductory course." He called for all in the profession to "experiment and evaluate, question everything, and take nothing for granted."

Cobb also noted that change is hard: "changing curriculum, like moving a graveyard, depends on local conditions." We concur, and reiterate the importance of faculty development efforts and resources created by organizations like the American Statistical Association and the Consortium for the Advancement of Undergraduate Statistics Education (CAUSE) to support instructors and the profession.

The revised GAISE College Report provides guidance for next steps along that path. To be realized, extensive faculty development and training is necessary to help instructors achieve the new learning outcomes that are emphasized in this report and to highlight the importance of teaching statistics in the spirit of the GAISE recommendations. Rob Gould suggests in his *International Statistical Review* article that more examples that resonate with modern students are needed. Additional efforts are needed to bridge the gap between statistical education and statistical practice. Such initiatives will help ensure that the introductory statistics course remains a vibrant choice for students seeking to learn how meaning can be extracted from the data around them.

## Further Reading

Cobb, G.W. 2015. Mere renovation is too little too late: We need to rethink our undergraduate curriculum from the ground up. *The American Statistician* (69)4:266-282. http://amstat.tandfonline.com/doi/suppl/10.1080/00031305.2015.1093029

delMas, R., Garfield, J., Ooms, A., and Chance, B. 2007. Assessing students' conceptual understanding after a first course in statistics. *Statistics Education Research Journal* 6:28-58. http://iase-web.org/documents/SERJ/SERJ6(2)_delMas.pdf

Franklin, C., Hartlaub, B., Peck, R., Scheaffer, R., Thiel, D., and Freier, K. T. 2011. AP statistics: Building bridges between high school and college statistics education. *The American Statistician* 65:177-182. http://amstat.tandfonline.com/doi/abs/10.1198/tast.2011.09111.